\documentclass[10pt]{iopart}

\usepackage{graphicx}

\usepackage[decimalsymbol=., separate-uncertainty = true, multi-part-units=single, range-phrase=-]{siunitx}

\usepackage[colorinlistoftodos]{todonotes}

\usepackage{float}

\usepackage{cite}

\usepackage{xfrac} %for \sfrac

\usepackage{amssymb}

\usepackage{hyperref}
\hypersetup{colorlinks, linkcolor=blue, urlcolor=blue}  % make hyperlinks blue

\DeclareSIUnit\sccm{sccm}
\DeclareSIUnit\jansky{Jy}
\DeclareSIUnit{\sqrthz}{\ensuremath{\sqrt{\text{\hertz}}}}
\DeclareSIUnit{\sqrtsec}{\ensuremath{\sqrt{\text{\second}}}}
\DeclareSIUnit{\sq}{\ensuremath{\Box}}
\DeclareSIUnit\atmosphere{atm}
\DeclareSIUnit\inch{''}
\DeclareSIUnit\dBm{dBm}

\begin{document}

\title[]{Characterization of sputtered hafnium thin films for high quality factor microwave kinetic inductance detectors}

\author{G. Coiffard$^{1,*}$, M. Daal$^{1}$, N. Zobrist$^{1}$, N. Swimmer$^{1}$, S. Steiger$^{1}$, B. Bumble$^{2}$ and B. A. Mazin$^{1}$}

\address{$^{1}$Department of Physics, University of California, Santa Barbara, California 93106, USA}
\address{$^{2}$NASA Jet Propulsion Laboratory, Pasadena, California 91109, USA}

\ead{$^{*}$gcoiffard@ucsb.edu}
\vspace{10pt}
\begin{indented}
\item[]\today
\end{indented}

\begin{abstract}
Hafnium is an elemental superconductor which crystallizes in a hexagonal close packed structure, has a transition temperature T\textsubscript{C} $\simeq \SI{400}{\milli\kelvin}$, and has a high normal state resistivity around \SI{90}{\micro\ohm\cm}. In Microwave Kinetic Inductance Detectors (MKIDs), these properties are advantageous since they allow for creating detectors sensitive to optical and near infra-red radiation. In this work, we study how sputter conditions and especially the power applied to the target during the deposition, affect the hafnium T\textsubscript{C}, resistivity, stress, texture and preferred crystal orientation. We find that the position of the target with respect to the substrate strongly affects the orientation of the crystallites in the films and the internal quality factor, Q\textsubscript{i}, of MKIDs fabricated from the films. In particular, we demonstrate that a DC magnetron sputter deposition at a normal angle of incidence, low pressure, and low plasma power promotes the growth of compressive (002)-oriented films and that such films can be used to make high quality factor MKIDs with Q\textsubscript{i} up to 600,000.
\end{abstract}

%\noindent{\it Keywords}: microwave kinetic inductance detectors, hafnium, crystallography
\submitto{\SUST}
%\maketitle
\ioptwocol

\section{Introduction}
Microwave Kinetic Inductance Detectors (MKIDs) \cite{Nature} are superconducting detectors capable of measuring the arrival time and energy of single photons. Detection of individual photons from infrared \cite{ARCONS, DARKNESS:Seth, Szypryt:instruments} to x-rays \cite{Gerhard:xray} have been demonstrated using MKIDs. In these detectors, energy deposited onto the superconducting film breaks Cooper pairs into excited electrons (quasiparticles) whose presence affects the total kinetic energy of the super current. The superconducting film is patterned into a resonant circuit, and changes in the kinetic energy of the super current manifest as changes in surface inductance, L\textsubscript{s}, of the resonant circuit. The resonant circuit is probed near its resonant frequency and the phase and amplitude of the transmitted probe signal are monitored for changes as quasiparticles are created.

A small superconducting gap energy, $2\Delta_0$, and a large kinetic inductance fraction are needed to make sensitive, responsive MKIDs. A small gap allows more quasiparticles to be created when a photon hits the detector resulting in greater sensitivity. A large surface inductance, L\textsubscript{s} a material property related to the kinetic inductance via a geometric proportionality constant, is valuable since the response of the detector is proportional to the fraction of kinetic inductance to total inductance in the resonator.

According to the BCS theory of superconductivity\cite{BCS:theory}, the transition temperature of the superconductor, T\textsubscript{C}, is related to the gap energy as $2\Delta_0 = 3.52 k_B \text{T}_C$. Assuming that the film is thin compared to its penetration depth, it can be derived \cite{Zmuidzinas:Review} that

\begin{equation}
\text{L}_s = \frac{\hbar}{\pi}\frac{ \rho_n}{ \Delta_0 t} = \frac{\hbar}{\pi}\frac{ \text{R}_s}{ \Delta_0}
\label{eq:Ls}
\end{equation}

where $t$ the film thickness and $\rho_n$ ($\text{R}_s$) is the normal state resistivity (sheet resistance) just prior to the superconducting transition. Note that the easily measurable T\textsubscript{C} can be substituted for $\Delta_0$, assuming BCS superconductivity. Choosing a superconductor with a large $\text{L}_s \geq \SI[per-mode=symbol]{10}{\pico \henry \per \sq}$ (high $\rho_n$) and low $\text{T}_C \leq \SI{1}{\kelvin}$ (low gap) allows us to design lumped element resonators in the \SI{4}{}-\SI{8}{\GHz} band sensitive to optical and near infra-red photons.

Only a few superconductors are known to posses relatively low T\textsubscript{C}, relatively high $\rho_n$, and make high quality resonators. Examples include TiN\textsubscript{x} \cite{ARCONS, TiN_LeDuc}, PtSi\textsubscript{x} \cite{ DARKNESS:Seth} and granular aluminum \cite{GranularAl}.
Our interest in hafnium thin films stems from the fact that it has a low T\textsubscript{C} around \SI{400}{\milli \kelvin}, a high $\rho_n$, and, as we have shown in a previous paper, can create resonators with Q\textsubscript{i} up to \num{190000}\cite{Nick:Hafnium}. Moreover, a material with a larger L\textsubscript{s} can be made thicker for a given detector responsivity, which allows for more flexible resonators designs and could help to reduce noise due to phonon escape from the superconductor \cite{Kozorezov:phonon}. Additionally, as an elemental superconductor, we can reasonably expect to obtain highly uniform films across the wafer, in contrast to reactively sputtered TiN\textsubscript{x} where the nitrogen gas flow distribution affects the superconducting gap uniformity of the deposited material \cite{Vissers_TiN, TiN:Martinis}. PtSi\textsubscript{x} also has its disadvantages as Pt is an expensive material and the process requires a very precise control of the Pt and Si  deposition rates as well as a fine control of the in-situ annealing of the bi-layer.

In order to improve hafnium MKID array performance, the research we present here is focused on better understanding the material properties and deposition conditions best suited for the fabrication of high quality factor hafnium resonators. This work adds to the existing literature in hafnium thin films developed for superconducting tunnel junction detectors \cite{Hf-junction-Kim, Hf-junction-Kraft} and transition edge sensors \cite{Hf-TES}.

\begin{figure}
	\centering
	\includegraphics[width=0.4\linewidth]{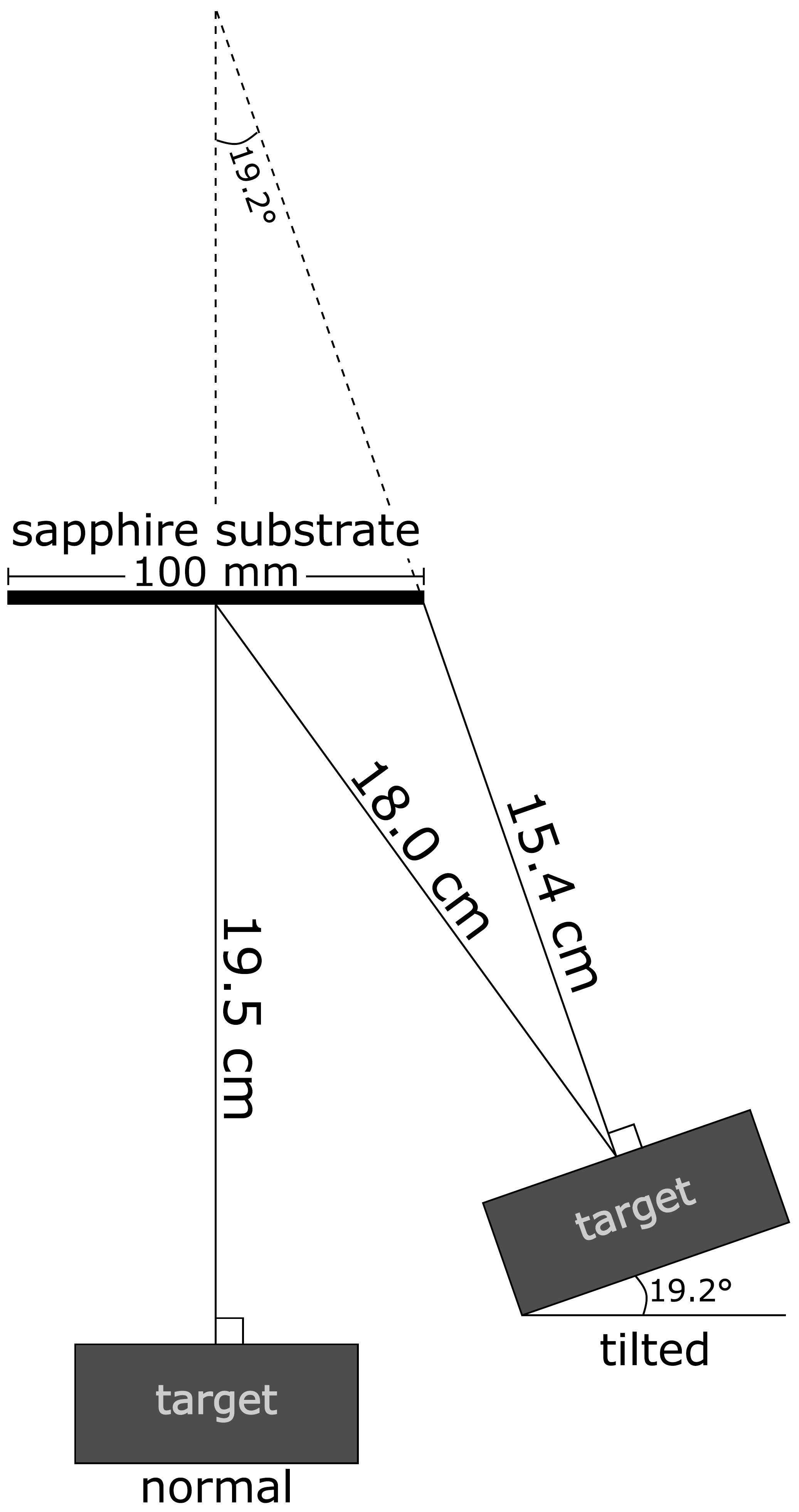}
	\caption{Diagram of the two different geometries of the sputter system. The hafnium target is either tilted by an angle of \SI{19.2}{\degree} off the perpendicular or at a normal incidence (\SI{90}{\degree}) geometry. The corresponding distance between the center of the \SI{75}{\mm} diameter target and the substrate are \SI{19.5}{\cm} and \SI{18.0}{\cm} respectively. In the tilted gun configuration, the distance between the center of the target and the edge of the substrate where the plume is aimed is \SI{15.4}{\cm}. All the distances are to scale.}
	\label{fig:systemgeometry}	
\end{figure}

\section{Hafnium Film Deposition}
%\subsection{Sputter deposition}
Hafnium films are deposited on \SI{100}{\mm} diameter a-plane sapphire substrates in a load-locked ultra-high vacuum AJA ATC-2200 sputtering system with a typical base pressure of \SI{6e-10}{\torr}. The argon working gas cylinder has a certified purity of 6N and it goes through an additional gas purifier before entering the sputter chamber. The hafnium sputter target used has a nominal purity of 3N5 (99.95 wt\%), but the actual ingot chemistry was measured to have 0.16 wt\% zirconium; additionally, impurities measuring above 1 ppm-wt are 55 ppm-wt oxygen, 16 ppm-wt carbon, 6.6 ppm-wt niobium, and 1.4 ppm-wt iron. Two different gun-substrate configurations were used for the hafnium film depositions (Figure \ref{fig:systemgeometry}). One where the sputter gun is tilted at an angle of \SI{19.2}{\degree} off normal and the distance between the center of the target and the closest point to it on the substrate is \SI{15.4}{\cm}. Another, where the target is concentric with  the substrate and there is \SI{19.5}{\cm} vertical distance between their centers. For each configuration, the substrate is rotating. All guns have a balanced magnet configuration and \SI{75}{\mm} diameter target. The deposition pressure and DC power are varied within the ranges \SI{1}{}-\SI{15}{\milli \torr} and \SI{20}{}-\SI{660}{\watt} respectively. Argon flow is kept at \SI{30}{sccm} for all depositions. The deposition rate for each condition is measured with a step profilometer and is used to deposit \SI{125}{\nm} thick hafnium films. This thickness gives the resonators on our fabrication mask the total inductance required to achieve their designed resonant frequency.

\begin{figure}
	\centering
	\includegraphics[width=\linewidth]{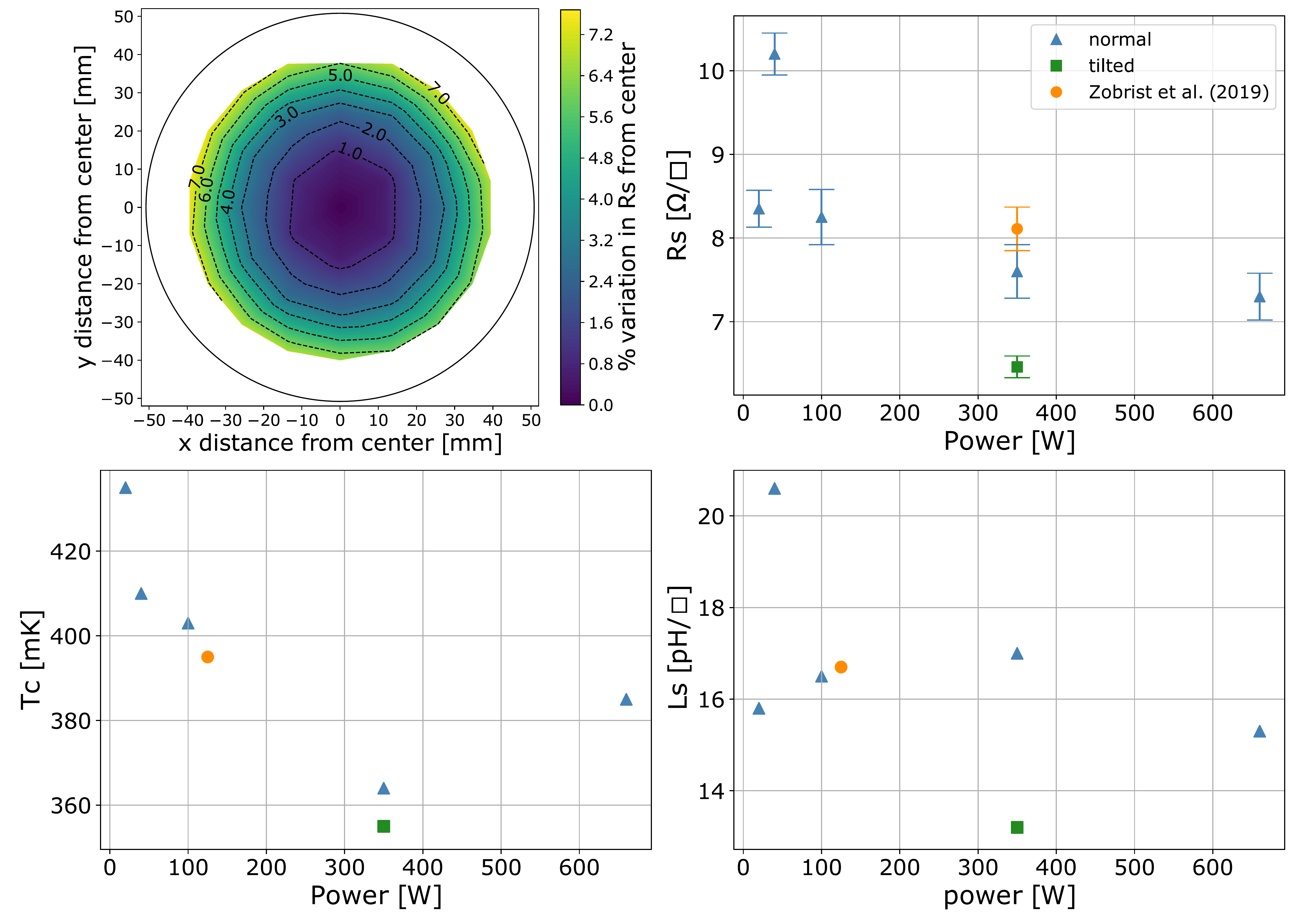}
	\caption{\textit{Top Left} The map shows the percent variation of the sheet resistance over a \SI{4}{\inch} wafer for a \SI{125}{\nm} thick hafnium film (\SI{4}{\inch} contour in black and wafer flat is facing up) deposited at \SI{2.5}{\milli \torr} and \SI{40}{\watt} and at a normal incidence. The average sheet resistance is \SI[per-mode=symbol]{10.20}{\ohm \per \sq}. The color scale shows the percent variation of the resistivity from the center of the wafer. \textit{Top Right} Sheet resistance for hafnium deposited at different power and angle. The error bars represents the standard deviation over the wafer. The data point from N. Zobrist et al. \cite{Nick:Hafnium} is also shown for comparison even though a different sputter system was used. The superconducting transition temperature (\textit{Bottom Left}) and calculated kinetic inductance using equation \ref{eq:Ls} (\textit{Bottom Right}) for the same films are also given.}
	\label{fig:resmapRsTcLs}
\end{figure}

\begin{figure*}
	%\centering
		\includegraphics[width=\linewidth]{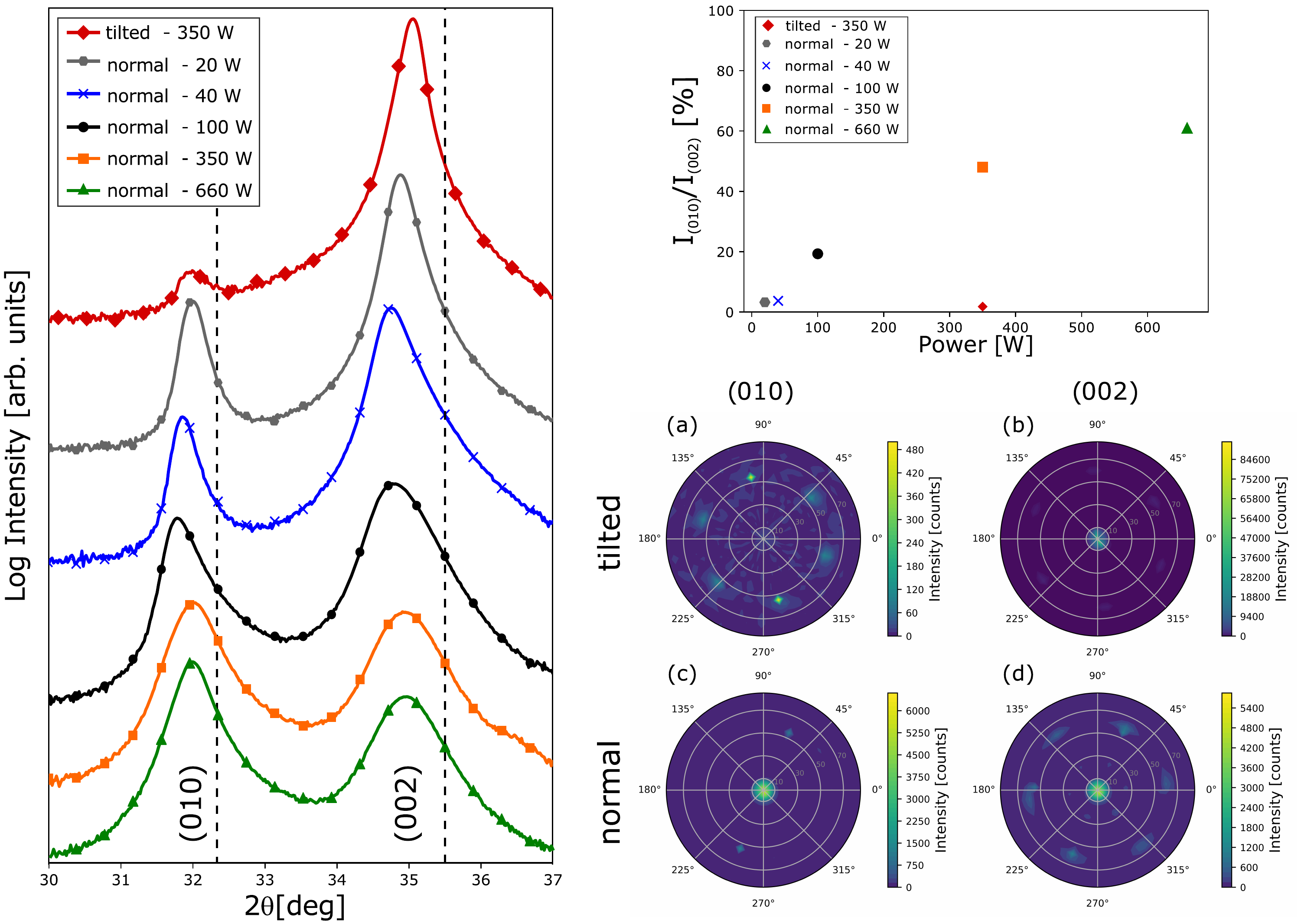}	
	\caption{\textit{left} $2\theta - \omega$ scans of \SI{125}{\nm} thick hafnium films on sapphire deposited at different gun tilts and  plasma powers. The patterns are shifted on the logarithmic intensity scale for clarity. \textit{top right} The ratio of intensity of the (010) peak to the (002) in percent as function of the deposition power is given. \textit{bottom right} (a) and (b) Pole Figure of (010) and (002) for the film deposited in the tilted gun configuration. (c) and (d) Pole figure of the (010) and (002) for the film deposited at \SI{660}{\watt} at a normal incidence. The intensity colormap is shown in a square root scale for clarity.}
	\label{fig:xray}
	\label{fig:polefig}
\end{figure*}

\section{Hafnium Film Characterization}
\emph{Superconducting transition temperature.} The superconducting transition temperature for each sample is measured in a Leiden Cryogenic dilution fridge with a base temperature of \SI{60}{\milli\kelvin}, and they are given in the Figure \ref{fig:resmapRsTcLs}. The T\textsubscript{C} of the films are found to be between \SI{435}{\milli\kelvin} and \SI{355}{\milli\kelvin} depending on the deposition conditions. We measure a residual-resistivity ratio (RRR), between room temperature and just above the superconducting transition temperature,  of 1.6 for the films deposited at a normal angle of incidence and 2.1 for the ones deposited at \SI{19.2}{\degree} off perpendicular (tilted configuration). The RRR value is used to connect the low temperature resistivity or sheet resistance to the value we measured at room temperature. All T\textsubscript{C} and RRR measurements are done on film samples taken from the center of the wafers.

\emph{Sheet resistance.} A CDE ResMap 4-point probe station mapper is used to measure the sheet resistance, $\text{R}_s$, of the hafnium films across the \SI{100}{\mm} substrates. From that data, we can quantify the film's uniformity and compute resistivities. Equation \ref{eq:Ls} implies that variations in the normal state sheet resistance just prior to transition accord to variations in gap parameter $\Delta_0$ (T\textsubscript{C}) and L\textsubscript{s} across the wafer. Gap parameter and L\textsubscript{s} variations result in pixel to pixel sensitivity and energy resolution variations as well as shifted resonant frequencies across the array. These effects result in a lower number of usable pixels. Consequently, obtaining superb sheet resistance uniformity is a fabrication priority.

Figure \ref{fig:resmapRsTcLs} presents a map of the sheet resistance we measure across a \SI{100}{\mm} wafer (the probe does not measure the resistance near the edge). The bulls-eye pattern is typical for all sputter conditions, though the uniformity we measured ranges from 7.0 \% in the tilted condition, to about 12.2 \% for the normal condition. Table \ref{tab:hafdata} summarizes the typical percent sheet resistance uniformity, $(\text{R}_{s,Max} - \text{R}_{s,Min})/{\text{R}_{s,Avg}}$, across the \SI{100}{\mm} wafer for each sputter condition. Our hafnium uniformity is comparable the  the uniformity we get for PtSi\textsubscript{x} \cite{Paul:PtSi}, and much better than for sub-stoichiometric TiN\textsubscript{x} which was never found to be better than \SI{\sim20}{\%} despite many optimizations in the deposition process  \cite{TiN_cristallo_NIST}. We believe that our sheet resistance uniformity is improvable by using a larger sputter target and positioning the substrate farther away from it during deposition, but it will be necessary to check that such changes do not negatively impact the resonator quality factors. 

The kinetic inductance of the films is calculated using equation \ref{eq:Ls} and the measured R\textsubscript{s} and T\textsubscript{C}; Figure \ref{fig:resmapRsTcLs} shows the value of the surface inductance for the different deposition conditions. It is found to be in the range \SI{15}{}-\SI[per-mode=symbol]{20}{\pico \henry \per \sq}.

\emph{Crystallography.} The crystallographic orientations and texture of the films are determined with a Panalytical XPERT MRD (Materials Research Diffractometer) PRO equipped with a Pixcel 3D X-ray diffractometer detector. The X-ray diffraction (XRD) patterns of \SI{125}{\nm} thick hafnium films grown on sapphire at 5 different plasma powers at a normal incidence angle are shown in Figure \ref{fig:xray}. A diffraction pattern for an hafnium film deposited with the tilted gun configuration is also given (in red). The analyses were made in the $2\theta - \omega$ mode between \SIrange{30}{37}{\degree}. The patterns show that the films are in the hexagonal phase \cite{hafnium-phase, Hf_crystal_McMurdie} and are strongly oriented along the (010), m-plane, and (002), c-plane, axes since only those two peaks are visible (no other peaks are visible when the analysis is performed on the full \SI{20}{}-\SI{80}{\degree} except the one from the sapphire substrate). The two peaks are shifted toward lower angles from the theoretical peak position (vertical dashed lines at \SI{32.3}{\degree} and \SI{35.5}{\degree}) which indicates that the films are under compression. Assuming our hafnium is predominantly polycrystalline and only oriented along two axes, the fraction of (010) and (002) crystal in the film can be computed by comparing the intensities of the peaks. Figure \ref{fig:xray} shows the percentage of (010)-oriented crystals in the film as a function of the plasma power used for the deposition. We show that by changing the plasma power during the deposition we are able to control the preferred orientation of the hafnium from a 3\% (100)-oriented film at \SI{20}{\watt} to 61.1\% at \SI{660}{\watt}. 

Assuming the diffraction peaks are only broadened due to the size of the crystallites, the Scherrer formula, $D=\frac{K \lambda}{\beta \cos(\theta)}$ \cite{ScherrerFormula}, can be used to estimate their sizes. $D$ is the size of the crystallite, $K$ is the shape factor and typically equals $0.9$, $\lambda = \SI{0.1546}{\nm}$ is the x-ray source wavelength, $\beta$ is the full width at half maximum of the peak in radian and $\theta$ is the Bragg angle in radian. For the films deposited at a normal incidence, the crystallites' dimensions range from \SI{12}{\nm} to \SI{30}{\nm}, and their dimensions are about \SI{40}{\nm} for the hafnium deposited at an angle of \SI{19.2}{\degree}.

\begin{figure}
	\centering
	\includegraphics[width=\linewidth]{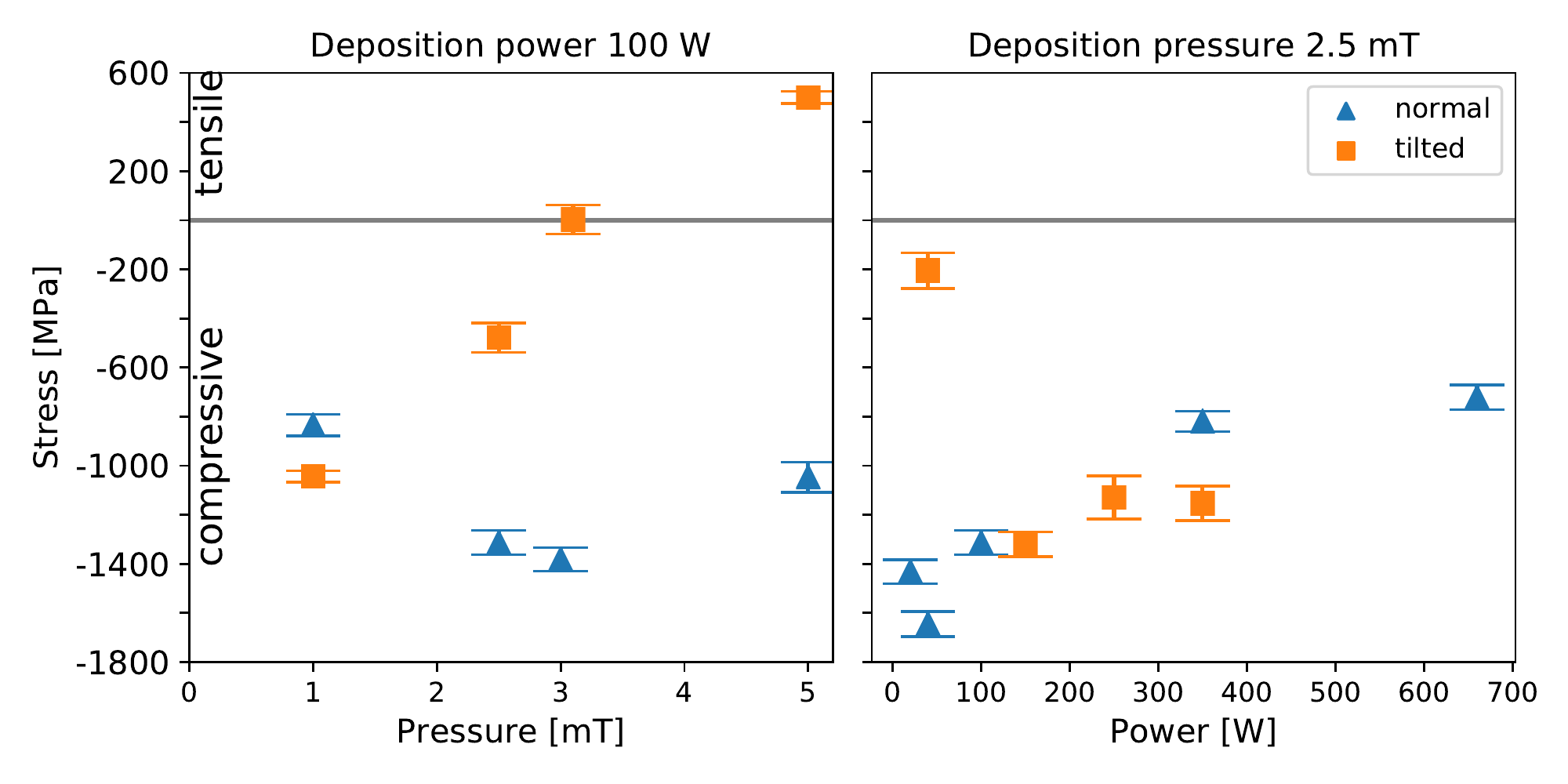}
	\caption{Average stress in the hafnium films deposited on a \SI{4}{\inch} wafer for the two deposition configurations. \textit{Left} At a fixed deposition plasma power of \SI{100}{\watt} and within the \SI{1}{}-\SI{5}{\milli \torr} pressure range. \textit{Right} At a deposition pressure of \SI{2.5}{\milli \torr} and for deposition power between \SI{20}{\watt} and \SI{660}{\watt}. The error bars represent the standard deviation of the measurement.}
	\label{fig:stresspressurepower}	
\end{figure}

We performed a texture measurement to understand the rotational orientation of the crystallites, which cannot be ascertained by the $2\theta - \omega$ peaks alone. In a pole figure, the intensity of the signal is proportional to the number of grains oriented along a particular direction, and a peak at the center corresponds to crystal oriented perpendicular to the surface of the film. The very weak signal in the center of the (010) pole figure, Figure \ref{fig:polefig} (a), shows that for the hafnium deposited at a gun angle of \SI{19.2}{\degree} the crystal planes are not growing perpendicular to the substrate but tilted by \SI{55}{\degree}. The (002) pole figure, Figure \ref{fig:polefig} (b), exhibits a strong peak at the center meaning that the crystallites are growing perpendicular to the substrate. In the case of the film deposited at a normal incidence of \SI{90}{\degree}, Figure \ref{fig:polefig} (c) and (d), the crystallites are mostly perpendicular to the substrate even though a fraction of them are tilted by \SI{55}{\degree}.

\emph{film stress.} The average stress of a \SI{125}{\nm} thick hafnium film deposited on a \SI{4}{\inch} diameter substrate is determined by measuring the change in radius of curvature prior to and after deposition by using a Tencor Flexus tool. Figure \ref{fig:stresspressurepower} shows the stresses obtained for a wide range of deposition conditions.
At a fixed deposition plasma power of \SI{100}{\watt} and for the tilted gun geometry, the stress changed from compressive to tensile when the pressure is increased. At a normal angle of incidence, only compressive films can be deposited within the \SI{1}{}-\SI{5}{\milli \torr} pressure range.
The stress for the films deposited at \SI{2.5}{\milli \torr} and a power between \SI{20}{} and \SI{660}{\watt} are always found to be compressive. In the case of the films deposited at a \SI{90}{\degree} angle, the stress in the films is between \SI{-800}{} and \SI{-1700}{\MPa}. 
Compressive hafnium films have been previously been studied by, Turner et al. \cite{Hafnium-Turner}, and have been used for the fabrication of the high performance MKID array by Zobrist et al. \cite{Nick:Hafnium}.
  
\begin{figure}
	\centering
	\begin{tabular}{cc}
		\includegraphics[width=0.45\linewidth]{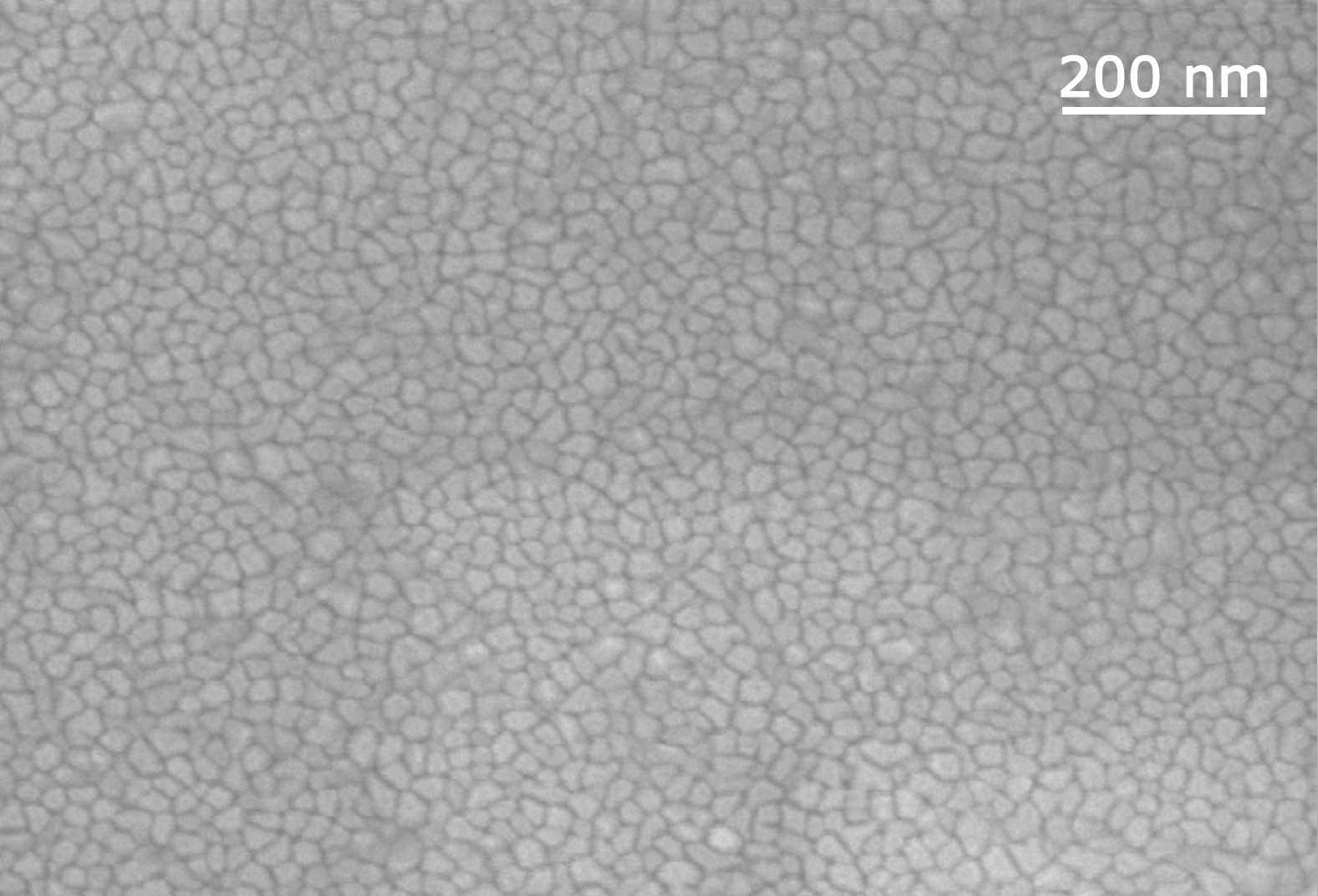} & \includegraphics[width=0.45\linewidth]{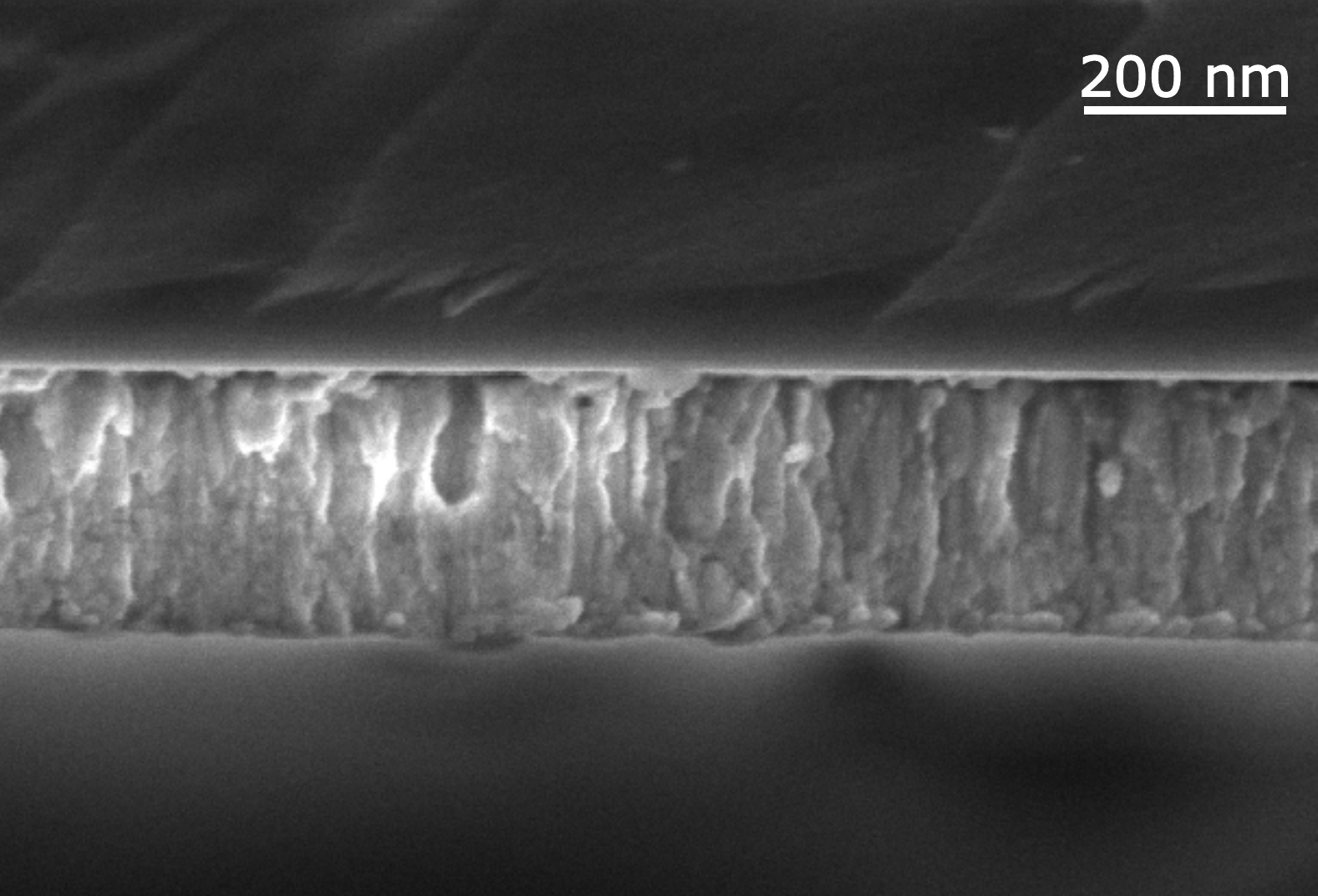} \\
		\includegraphics[width=0.45\linewidth]{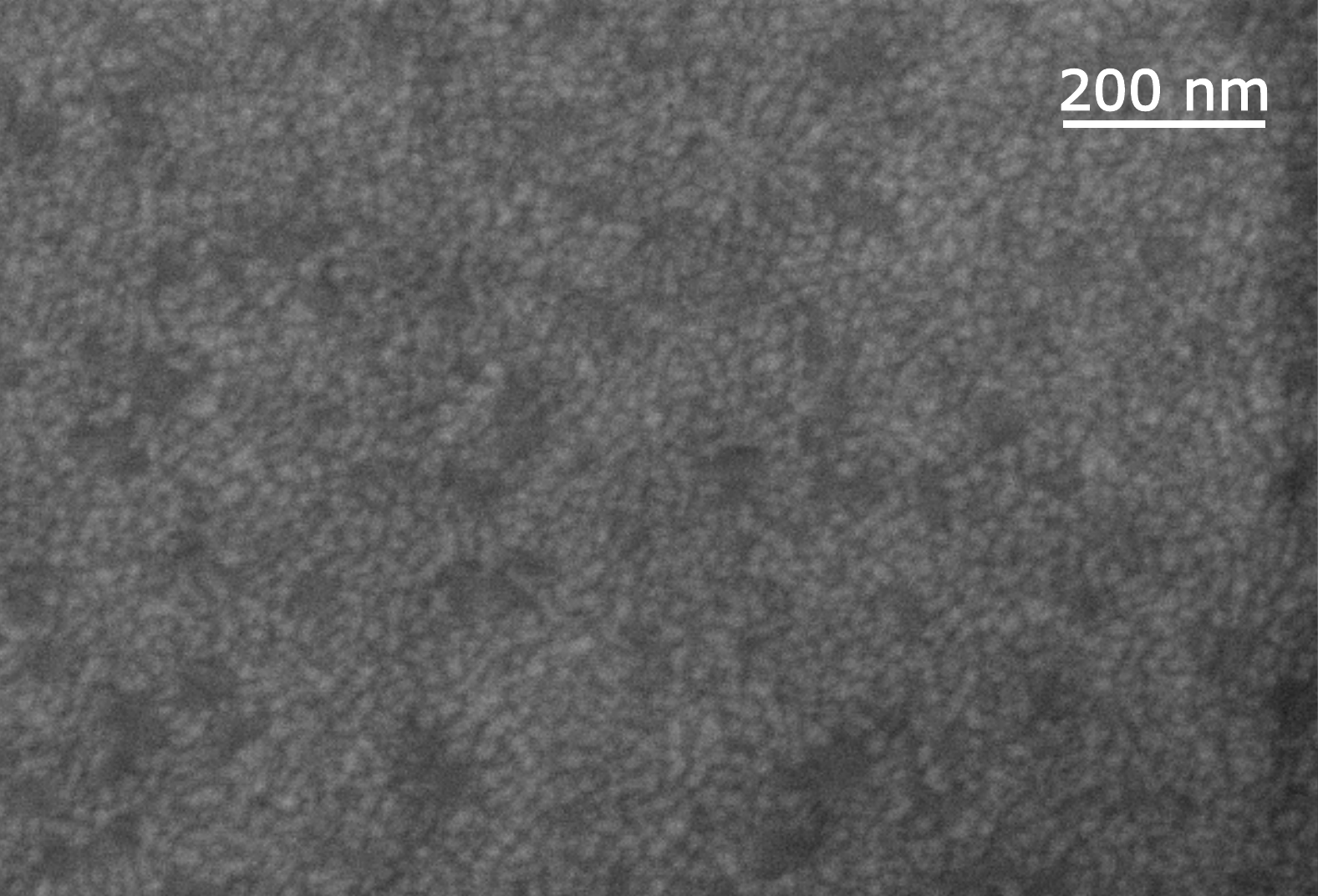} & \includegraphics[width=0.45\linewidth]{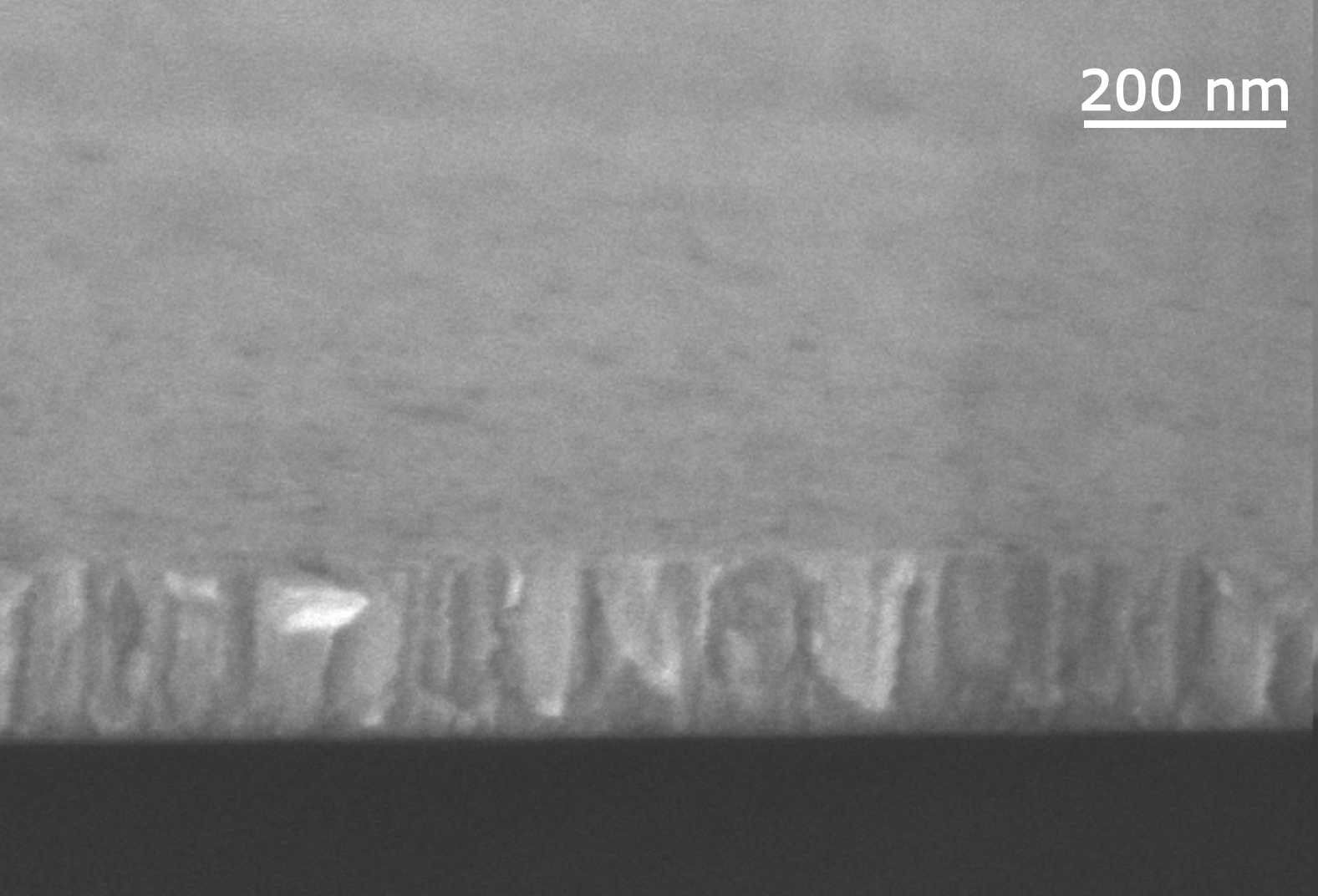}\\
	\end{tabular}

	\caption{SEM images of the surface and cross section of hafnium films deposited at \SI{350}{\watt} and \SI{2.5}{\milli \torr} with a gun tilt of \SI{19.2}{\degree} (\textit{top}) and with a normal angle of incidence (\textit{bottom}) The dark patches on the surface of the normal configuration are unidentified.}
	\label{fig:sem}
\end{figure}

At low deposition pressures or short target-to-substrate distance, sputtered atoms impinge upon the substrate without having lost energy to scattering events. The impinging atoms can have enough energy to create local defects as a result of the shockwave created by their impact in a phenomena known as peening. Peening provides a mechanism for the growth of dense, compressive films  \cite{DHeurle:stress, Windischmann:stress,Koch:stress}. While we have not measured the density of our films, their tendency to have giga-pascal scale compressive stress, our sputter pressure, and the target-substrate geometry lead us to suspect that peening plays a dominant role in the growth of our films.

Curiously, we find that the stress of our films as a function of the sputter power follows two different trends depending on the sputter gun configuration. Increasing the powers gives decreasing stress for the normal incidence configuration while the opposite is true for the tilted gun configuration. Meanwhile, we find that the only way to obtain tensile films is to increase the deposition pressure. We expect that in this situation the impinging atoms reach the substrate with  an insufficient energy to move to sites forming a closely packed atomic arrangement resulting instead in a porous, void-rich microstructure \cite{stress:Ta}.

\begin{figure}
	\centering
	\includegraphics[width=0.9\linewidth]{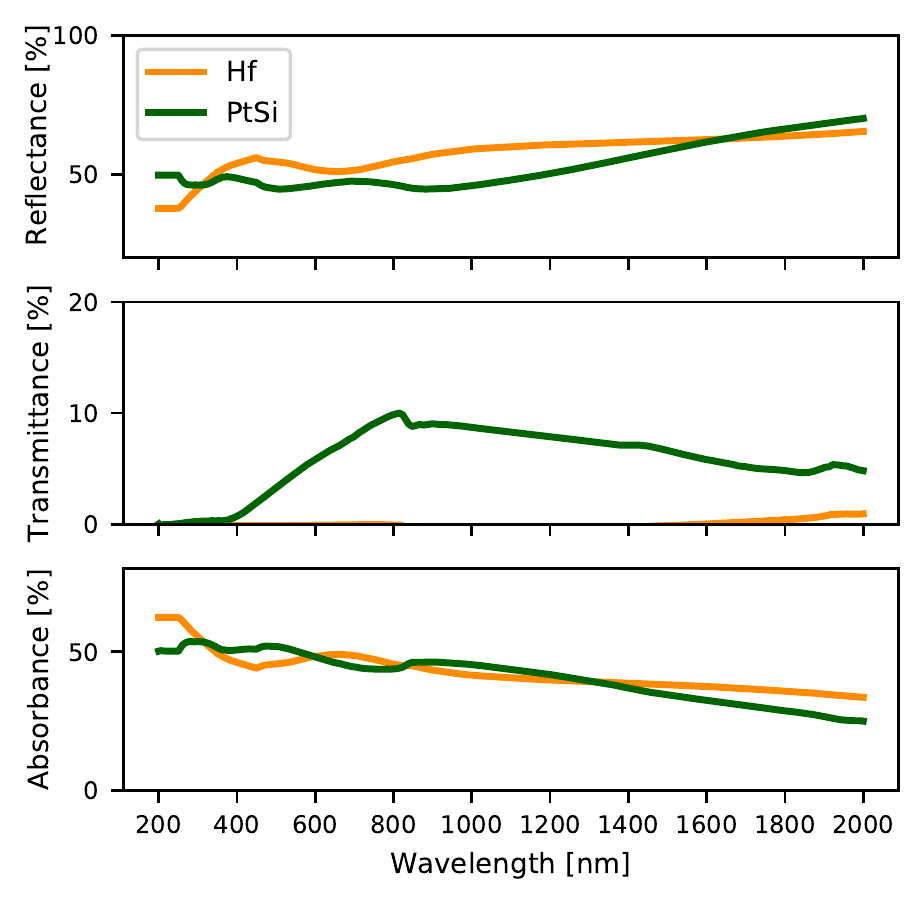}

	\caption{Optical reflectance, transmittance and absorbance measurement of a \SI{245}{\nm} hafnium and \SI{60}{\nm} PtSi\textsubscript{x} films deposited on sapphire substrates. The PtSi\textsubscript{x} spectra are given for comparison. The discontinuity at \SI{800}{\nm} comes from a change of the light source inside the spectrometer}

	\label{fig:opticaldata}	
\end{figure}

\emph{Film Morphology.} The surface and cross section of hafnium films are shown in Figure \ref{fig:sem}. Columnar growth is observed in all films and the crystallites are contained within the columns. Measuring from the SEM images, the crystalline grains are approximately \SI{32}{\nm} for the tilted gun sputter configuration and \SI{17}{\nm} for the normal configuration. This accords with the estimate based on the Sherrer formula. 

We also remark that the hafnium films readily charge up when imaged in the SEM. This is unusual for a metal. The film deposited at \SI{90}{\degree} shows some dark areas on the surface that have not been identified.

\begin{table*}
	\begin{center}
		%tableau un peu plus aéré
		\renewcommand{\arraystretch}{1.5}
		\setlength{\tabcolsep}{8pt}
		\begin{tabular}{c  c  c  c  c  c  c  c}
			\hline
			Target Power  &  R\textsubscript{s} & R\textsubscript{s} & T\textsubscript{C} & L\textsubscript{s} & $\frac{I_{(010)}}{I_{(002)}}$  & $\sigma$  & Q\textsubscript{i} \\
			
			[\SI{}{\watt}] &  [\SI[per-mode=symbol]{}{\ohm \per \sq}] &  uniformity & [\SI{}{\milli \kelvin}] & [\SI[per-mode=symbol]{}{\pico \henry \per \sq}] & \% & [\SI{}{\mega \pascal}]  &   \\
			
			\hline
			\hline
			\multicolumn{8}{c}{Normal configuration} \\
			
			\hline
			
			660 & $7.30 \pm 0.28 $ & 11.6 \% & 385 & 15.3 &  61.1 \% &$-722 \pm 50$ & $\num{325000} \pm \num{30000} $ \\
			
			350 & $7.60 \pm 0.32 $ & 12.2 \% & 364 & 17.0 &  48.7 \% & $-820 \pm 42$ & $\num{174000} \pm \num{17000} $  \\
			
			100 & $8.25 \pm 0.33 $ & 11.3 \% & 403 & 16.5 &  19.3 \% & $-1314 \pm 50$ & $\num{405000} \pm \num{22000} $ \\
			
			40 & $10.20 \pm 0.25 $ & 7.3 \% & 410 & 20.6 &  3.7 \% & $-1408 \pm 51$ &  $\num{605000} \pm  \num{80000} $ \\
			
			20 & $8.35 \pm 0.22 $ & 8.1 \% & 435 & 15.8 &  3.0 \% &$-1432 \pm 48$ &   $\num{515000} \pm \num{95000} $ \\
			\hline
			\multicolumn{8}{c}{Tilted configuration} \\
			\hline
			\hline			
			350 & $6.46 \pm 0.13 $ & 7.0 \% & 355 & 13.2 & 2.0 \% & $-1154 \pm 71$ & $\num{16800} \pm \num{2000}$ \\
			\hline
			\multicolumn{8}{c}{Normal configuration - Zobrist et al. \cite{Nick:Hafnium}} \\
			\hline
			\hline			
			125$^*$ & $8.11 \pm 0.26 $ & 12.1 \% & 395 & 16.7 & 38 \% & $-1236$ & $\num{190000}$ \\
			\hline
		\end{tabular}

		\caption{Summary of the properties of \SI{125}{\nm} hafnium films deposited at \SI{2.5}{\milli \torr} argon pressure, with different plasma powers and gun angles. Resistivity (with standard deviation over the wafer) and T\textsubscript{C} ($\pm \SI{5}{\milli \kelvin}$ thermometer calibration) at the center of the wafer, preferred crystal orientation and stress (with standard deviation over the wafer) in the films. The average and standard deviation of the internal quality factor Q\textsubscript{i} for MKIDs made out of the specific films are given. The device studied in Zobrist et al. \cite{Nick:Hafnium} was fabricated at JPL. That hafnium was DC magnetron sputtered at normal incidence using a \SI{6}{\inch} target at \SI{2.5}{\milli \torr}. $^*$We scaled the \SI{350}{\watt} power used on the \SI{6}{\inch} target to what it would be on a \SI{3}{\inch} target keeping power density fixed and it gives a power of \SI{125}{\watt}.}

		\label{tab:hafdata}
	\end{center}
\end{table*}

\emph{Optical Film Properties.} The room temperature optical reflectance, transmittance and absorbance have been measured within the \SIrange{200}{2000}{\nm} range with a Shimadzu UV3600 spectrometer equipped with an integrating sphere. The spectra are plotted in Figure \ref{fig:opticaldata}. This room temperature measurement allows us to set an upper limit to the quantum efficiency of the material for use in detectors sensitive to these wavelengths. Because these measurements are made at room temperature, we adjust the film thicknesses to match the resistivity of a \SI{125}{\nm} thick film just above the superconducting transition temperature. Using RRRs of 1.6 and 1 for hafnium and PtSi\textsubscript{x} the spectra are acquired from films \SI{245}{\nm} and \SI{60}{\nm} thick, respectively. We compare Hf to PtSi\textsubscript{x} because we have previously established that PtSi\textsubscript{x} is suitable for making high quantum efficiency MKIDs sensors \cite{DARKNESS:Seth, Szypryt:instruments}. We find that the  Hf has a lower transmittance than PtSi\textsubscript{x}, but they have similar reflectance and absorbance.

\section{Resonator Fabrication}
To make the MKIDs, \SI{125}{\nm} thick Hf films first undergo a dehydration bake at \SI{135}{\degreeCelsius}. We then spin on a \SI{80}{\nm} thick layer of DUV-42P6 adhesion promoter followed by a \SI{800}{\nm} thick layer of UV6-0.8 imaging photoresist. The resist is patterned with and MKID test geometry using an ASML PAS 5500/300 DUV stepper which has a resolution better than \SI{200}{\nm} for dense patterns. The resist is developed in AZ MIF 300 and the adhesion promoter layer is removed in the etch chamber as part of hafnium etch. The hafnium is etched in a PlasmaTherm SLT 700 reactive ion etcher (RIE) which has \SI{4}{\inch} parallel plates and water cooling to \SI{20}{\celsius} for the substrate. (We note that wafer temperature never exceeded \SI{41}{\degreeCelsius} during the etch and deposition for all deposition conditions.) The etch recipe is BCl\textsubscript{3} and Cl\textsubscript{2} flowed at \SI{60}{\sccm} and \SI{40}{\sccm}, respectively; \SI{5}{\milli \torr} process pressure; \SI{100}{\watt} RF power at \SI{13.56}{\MHz}. The etch rate was measured with a step profilometer and is found to be \SI[per-mode = symbol]{0.41}{\nm \per \second} and uniform over a full \SI{4}{\inch} wafer. The resist is then removed with solvents, gold bond pads are added (via a lift-off process) on the side of the chip to ensure a good thermalization of the devices to the bath temperature and finally diced. The chip dimensions are $\SI{13}{} \times \SI{13}{\mm}$.

\begin{figure}
	\centering
	\includegraphics[width=\linewidth]{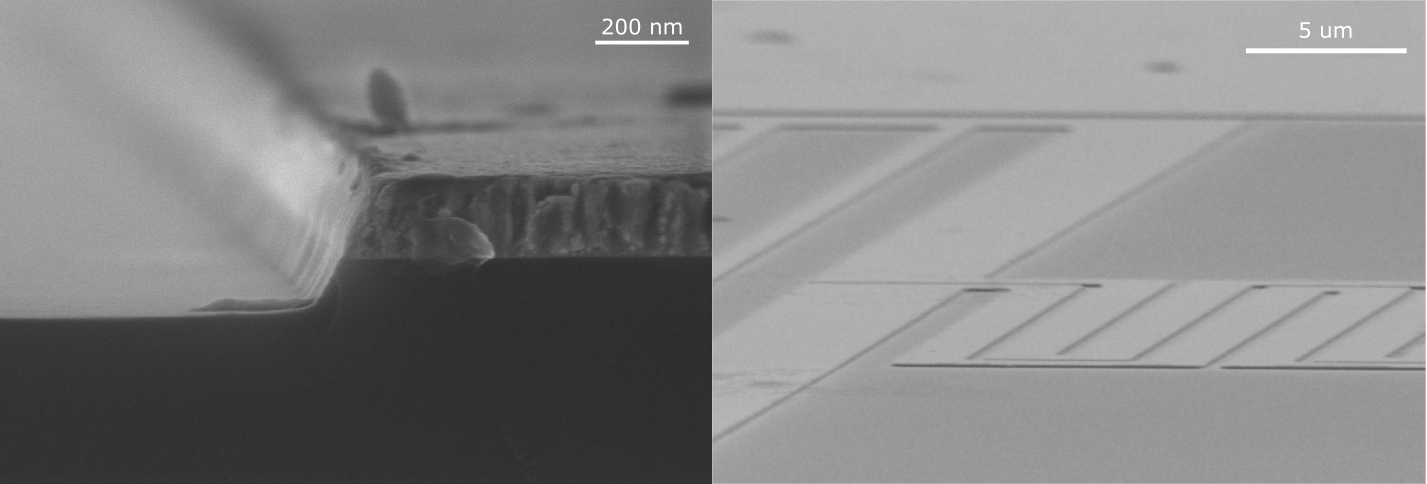}
	\caption{SEM images of the profile of a hafnium film etched by reactive ion etching in a mixture of BCl\textsubscript{3}/Cl\textsubscript{2}.}
	\label{fig:semetch}
\end{figure}

The etch profile is an important parameter to control to make MKID devices. The smallest feature on our lumped element test geometry is the gap between the meandered inductor trace which is \SI{500}{\nm} wide. Vertical walls are needed to avoid shorts. Figure \ref{fig:semetch} shows SEM images of the etch profile of an hafnium film deposited on a silicon substrate. Our etch process produces sidewalls that are approximately \SI{110}{\degree}. Occasional etch residue, which degrades MKID performance \cite{etch_TiN}, is observed in the etch trenches and on the etch side wall.

The resonators are designed to have resonant frequencies around \SI{5}{\GHz} and to have a spacing of \SI{2}{\MHz}. In order to compare the performances of each chip, nine resonators with a simulated coupling quality factor Q\textsubscript{C} of \num{40000} are analyzed.

\section{Resonator Measurements}
\label{section:measurement} 
The chips are cooled down in a BlueFors dilution refrigerator to a base temperature of $ \SI{\sim12}{\milli \kelvin}$. Resonators are first found manually with an Agilent \SI{20}{\GHz} vector network analyzer. Then, an automated readout system which uses a synthesizer and digitizer is used to find the power at which each resonator starts to bifurcate \cite{Jiansong:PhD}. This system then acquires the complex IQ transmission data by sweeping the microwave probe through the resonance from low to high frequency using a probe power that is \SI{14}{\dBm} below bifurcation. The resulting IQ curves are fitted with the model proposed by Khalil et al. \cite{Khalil:fits}. From the fitted curves, the intrinsic quality factor of each resonator is computed. 

\begin{figure}
	\centering
	\includegraphics[width=\linewidth]{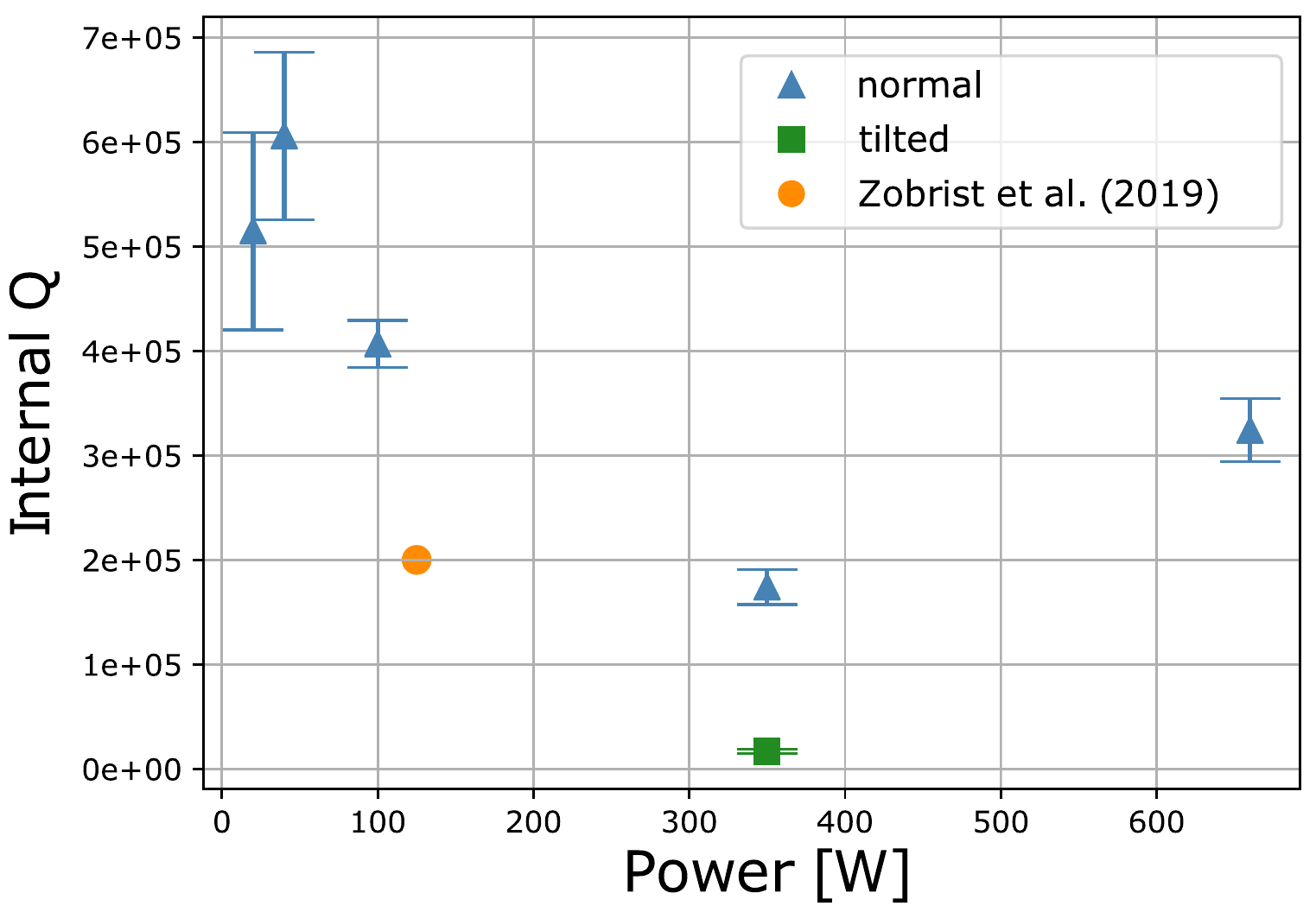}
	\caption{Except for the Zobrist et al.  point, each data point is the average internal quality factor from nine resonators on a single test chip made from a Hf film deposited at the power indicated on the x axis. From chip to chip, the same nine resonators are included in the average. The quality factor for the film deposited at an angle of \SI{19.2}{\degree} is also shown. The error bars represent the statistical standard deviation of the internal quality factor of the resonators. The Zobrist et al. point comes from the average over 10 resonators (also around \SI{5}{\GHz}) of a slightly different resonator geometry from a different MKID test chip mask.}
	\label{fig:QivsPower}
\end{figure}

An MKID test chip was fabricated out of the Hf film produced by each sputter condition. In Figure \ref{fig:QivsPower}, we plot the average Q\textsubscript{i} of the nine resonators described in the previous section for these sputter conditions. All the hafnium MKIDs made from films deposited in the tilted gun configuration gave low quality resonators with $\text{Q}_i < \num{17000}$. We are able to obtain Q\textsubscript{i} as high as \num{600000} in the normal incidence gun configuration, at low power. 

We find that our hafnium resonators have Q\textsubscript{i} on par with that measured from PtSi\textsubscript{x} \cite{Paul:PtSi}. The same one layer MKID test geometry was used for both the  PtSi\textsubscript{x} and Hf film resonators. 

\section{Conclusions}In conclusion, we have studied the properties of sputtered hafnium films and MKIDs made from them while varying several sputter deposition parameters.
We have shown that we are able to control the crystal orientation by growth conditions. Sputtered hafnium grows along the (010), m-plane, and (002), c-plane, crystal planes, and deposition at a normal angle of incidence promotes crystal growth perpendicular to the substrate. By decreasing the deposition power, the film tends to be mostly (002)-oriented. 

When the hafnium is sputtered at a normal angle of incidence to the substrate, we are able to make MKID resonators with internal quality factor (Q\textsubscript{i}) as high as \num{600000}. We have observed this effect in two different sputter systems. Such Q\textsubscript{i} is about a factor of 3 higher than the Q\textsubscript{i}s achieved in our previous hafnium MKID paper \cite{Nick:Hafnium}. We have shown that resonator Q\textsubscript{i} increases with decreasing the sputter power; this coincides with an increase in the (002) plane to (010) ratio, compressive film stress, and smaller crystals grains as measured by scanning electron microscope.
When the hafnium is deposited at a gun angle of \SI{19.2}{\degree}, \num{17000} is the highest resonator Q\textsubscript{i} we can obtain.

The high Q\textsubscript{i} resonators have a superconducting transition temperature of \SI{435}{\milli \kelvin} and a surface inductance of \SI[per-mode=symbol]{16}{\pico \henry \per \sq}. From our optical reflectance measurement, we expect the quantum efficiency of sputtered Hf MKIDs for the detection of photons in the UVOIR band to be on par with PtSi\textsubscript{x} resonators \cite{Paul:PtSi} which we currently use in our detector arrays \cite{DARKNESS:Seth, Szypryt:instruments}. The sheet resistance uniformity obtained with hafnium is in line with PtSi\textsubscript{x} and routinely better than TiN\textsubscript{x} and methods for improving it are clear, e.g., using a larger target and placing the substrate farther away from it. 

Our research has not determined why a certain combinations of sputter gun angle, pressure, and power yield to high Q\textsubscript{i} MKID resonators, but we have identified film characteristics, such as crystal structure, texture, film stress, and grain size which correlate with high Q\textsubscript{i}.

\section*{Acknowledgment}
This work was supported by the NASA APRA program under grant NNX16AE98G. N.Z. was supported throughout this work by a NASA Space Technology Research Fellowship. The research reported here made use of shared facilities of the UCSB MRSEC (NSF DMR 1720256), a member of the Materials Research Facilities Network (\url{www.mrfn.org}). The author would like to thank Youli Li from the UCSB Material Research Laboratory for his great help with the X-ray diffraction measurements and analysis. 

\section*{References}
\bibliography{biblioGreg}

\end{document}